\documentclass[lettersize,journal]{IEEEtran}
\usepackage{amsmath,amsfonts}
\usepackage{algorithmic}
\usepackage{algorithm}
\usepackage{array}
\usepackage[caption=false,font=normalsize,labelfont=sf,textfont=sf]{subfig}
\usepackage{textcomp}
\usepackage{stfloats}
\usepackage{url}
\usepackage{verbatim}
\usepackage{graphicx}
\usepackage{cite}
\usepackage{multirow}
\usepackage{amssymb}
\usepackage{upgreek}
\usepackage{tikz,xcolor}
\usepackage{hyperref}
\hypersetup{hidelinks,
	colorlinks=true,
	allcolors=black,
	pdfstartview=Fit,
	breaklinks=true
}

\definecolor{lime}{HTML}{A6CE39}
\DeclareRobustCommand{\orcidicon}{
	\begin{tikzpicture}
		\draw[lime, fill=lime] (0,0)
		circle[radius=0.16]
		node[white]{{\fontfamily{qag}\selectfont \tiny \.{I}D}};
	\end{tikzpicture}
	\hspace{-2mm}
}
\foreach \x in {A, ..., Z}{%
	\expandafter\xdef\csname orcid\x\endcsname{\noexpand\href{https://orcid.org/\csname orcidauthor\x\endcsname}{\noexpand\orcidicon}}
}

\begin{document}

\title{A Dynamic Capacitance Matching (DCM)-based Current Response Algorithm for Signal Line RC Network}

\author{Zhoujie Wu\hspace{-1.5mm}\orcidA{}, Cai Luo\hspace{-1.5mm}\orcidB{} and Zhong Guan\hspace{-1.5mm}\orcidC{}
\thanks{Zhoujie Wu, Cai Luo and Zhong Guan are with the School of Microelectronics Science and Technology, Sun Yat-Sen University, Zhuhai, 519082, China. Email: \{wuzhj53, luoc23\}@mail2.sysu.edu.cn, guanzh23@mail.sysu.edu.cn}
\thanks{Corresponding Author: Zhong Guan.}}

\markboth{Journal of \LaTeX\ Class Files,~Vol.~14, No.~8, August~2021}%
{Shell \MakeLowercase{\textit{et al.}}: A Sample Article Using IEEEtran.cls for IEEE Journals}


\maketitle
\begin{abstract}
This paper proposes a dynamic capacitance matching (DCM)-based RC current response algorithm for calculating the current waveform of a signal line without performing SPICE simulation. Specifically, unlike previous method such as CCS model, driver linear representation, waveform functional fitting or equivalent load capacitance, our algorithm does not rely on fixed reduced model of both standard cell driver and RC load. Instead, our algorithm approaches the current waveform dynamically by computing current responses of the target driver for various load scenarios. Besides, we creatively use symbolic expression to combine the y-parameter of RC network with the pre-characterized driver library in order to perform capacitance matching by considering over/under-shoot effect. Our algorithm is experimentally verified on 40nm CMOS technology and has been partially adopted by latest commercial tool for other nodes. Experimental results show that our algorithm has excellent resolution and promising efficiency compared with traditional methods and SPICE golden result, especially for application in computing delay, power and signal line electromigration.
\end{abstract}


\section{INTRODUCTION}
\IEEEPARstart{I}{n} advanced technology, the parasitic effect of interconnect lines leads to an increasingly large scale RC circuit. It is necessary for electronic design automation (EDA) tools to efficiently and accurately complete complex design tasks. The RC current response of signal lines serves as a crucial parameter for various analyses, including timing, power, and signal line electromigration (EM) reliability analysis. While transistor-level simulators can provide results of gold-standard precision, their use in modern IC design is hindered by memory and time constraints. Consequently, it is common practice to abstract gate-level circuits and simplify the analysis through modeling.

Signal lines, which behave as RC circuits, become progressively more nonlinear as their aspect ratio and length increase, posing analytical challenges. For current estimation in RC circuits, several well-known algorithms based on moment matching such as and AWE \cite{ref1} and PRIMA \cite{ref2} have been used for decades. However, they require a known input either in a form of applied current or voltage waveform to predict the current response of the RC circuit. In the case of gate-level driving, this input also depends on the gate slew, making these algorithms less efficient, so the need to model the gate cell is unavoidable. Methods for timing modeling of gate cells can be broadly categorized as follows.

One approach is the Current Source Model (CSM). Criox and Wong introduced a gate cell current source model called Blade \cite{ref3}, comprising a voltage-controlled current source, internal capacitance, and a one-step time-shift operation. This model effectively simulates the electrical behavior from the input side to the output side of gate cells. Kellor further improved model accuracy by introducing the KTV model, which considers Miller capacitance \cite{ref4}. Subsequently, nonlinear characteristics of capacitance parameters in CSM models have been considered by Li et al. \cite{ref5} and Fatemi et al. \cite{ref6}, incorporating input and output parasitic capacitances, Miller capacitance between them, and the output current source as functions of input and output voltages. The CSM model has been further developed in \cite{ref7,ref8,ref9,ref10,ref11,ref12} to address issues such as multi-port, sequential cells, and feedback loops. Currently, widely adopted industry methods such as CCS \cite{ref13} and ECSM \cite{ref14} involve the establishment of driver and receiver model for each cell. The lookup table (LUT) characterizes the behavior of the cell at different gate slew and output loads, and the current waveform in the LUT is selected by the effective capacitance, with two different input capacitances $C1$ and $C2$ used to model the nonlinear receiver input transistor capacitance and the Miller effect. These models are independent of the load, allowing them to handle scenarios like input nonlinearity distortions or even non-monotonic behavior caused by crosstalk.

Another approach is Voltage Response Models (VRM), such as Non-Linear Delay Models (NLDM), which utilizes transistor-level simulation and records the corresponding gate delay and output delay of each standard cell under various input slews and output loads, enabling timing estimation by interpolation/extrapolation of LUT. Since employing the total interconnect capacitance $C_{total}$ is overly pessimistic, considering the influence of interconnect resistance, using iterative \cite{ref15,ref16,ref17,ref18} or non-iterative methods \cite{ref19,ref20} can identify effective capacitance to enhance accuracy. The former necessitates iterative calculations of $C_{eff}$ until it converges, typically requiring 5 to 10 iterations for convergence, and in cases of unreasonable initial values, incurring significant CPU time overhead. The latter computes an effective capacitance in a single calculation but necessitates a closed-form expression, and although it performs well in delay analysis, it cannot precisely match the output waveform, with slew errors that could be as high as 15\% \cite{ref21}. Beyond these challenges, as technology scaling, the complexity of RC load makes it challenging to perfectly fit the response curve of the RC network, rendering two-piece output \cite{ref15} or two effective capacitance \cite{ref18} insufficient for an ideal fit \cite{ref22}.

Apart from these methods, there are approaches that utilize fitting functions to predict output waveform. Since only the driver's input and the topology of the RC circuit are known, the waveform of the driver's output (input of the RC network) is directly modeled as a parameter-based analytical function. For example, the double exponential function is usually applied to model the current responses \cite{ref23}, while Weibull \cite{ref24} and gamma \cite{ref25} function are often used to model the voltage responses, as described in (\ref{eq1})-(\ref{eq3}). The related parameters are trained by minimizing the error for specific responses such as the response of the driver loaded with a fixed capacitance. Recently, a macromodeling approach based on the inertial delayed Elmore delay (DED) was proposed for fitting gate output in \cite{ref26}. This approach utilizes SPICE to extract two macromodel parameters of the gate cell under a single capacitance to rapidly approximate the delay and output waveform within an error of 5\%, but it fails to predict the initial over/under-shoot, a limitation that becomes more pronounced when the input slope is substantial.

\begin{equation}
	\label{eq1}
	I(t)=K\cdot (e^{-t/T_a}-e^{-t/T_b})
\end{equation}
\begin{equation}
	\label{eq2}
	V(t)=V_{DD}\cdot (1-exp((-\frac{t}{\beta})^\alpha))
\end{equation}
\begin{equation}
	\label{eq3}
	V(t)=V_{DD}\cdot (1-\frac{\Gamma (n,\lambda t)}{\Gamma (n)})
\end{equation}

Building upon our prior work \cite{ref27}, this paper proposes a novel method for solving RC load response waveform based on dynamic capacitance matching (DCM). It models RC load under the influence of logic gate as dynamic capacitance, employing symbolic expression to seamlessly integrate high-order driving point function of RC circuit with the driver pre-characterization library, calculating the values of dynamic capacitance for N segments while considering their interdependencies. Some approximations are used to ensure algorithmic stability. The algorithm predicts the current waveform by utilizing the current responses of fixed capacitances, either from foundry data or through pre-characterization using SPICE simulation if the data is unavailable. In comparison to traditional methods, even for extensive and intricate RC load, our algorithm can rapidly and accurately predict the nondigital behavior of signal lines. This technique excels in three critical aspects: 1) the degree of fit for voltage/current response curves; 2) computation time; and 3) configurability.

The remainder of this article is organized as follows. In Section \uppercase\expandafter{\romannumeral2}, we provide a relatively detailed explanation of the model order-reduction (MOR) used in our method. In Section \uppercase\expandafter{\romannumeral3}, we explain the theoretical basis of the proposed method and show the complete flow of the algorithm. In Section \uppercase\expandafter{\romannumeral4}, we compare the simulation results of the classical and the proposed methods. Finally, we conclude in Section  \uppercase\expandafter{\romannumeral5}.

\section{TYPICAL SIGNAL LINE RC MODEL}
Most signal lines can be modeled as pure RC networks. According to the characteristics of linear circuits, researchers are usually interested in their order reduction models. MOR is divided into two categories:
One is based on time domain. Typical methods include Chebyshev polynomials and Laguerre polynomials\cite{ref28,ref29,ref30,ref31}. A typical feature is that the state variables of the system are expanded by orthogonal polynomials in the time domain, and then the corresponding coefficient matrix of the orthogonal polynomials is used for projection order reduction.

The other is based on frequency domain technology and goes in two directions:

\underline{Moment Matching}. AWE\cite{ref1} is the first model applied to electronic circuits. By calculating the system moment explicitly and the coefficient of the transfer function of the reduced order system by means of moment matching, it can quickly analyze the reduced order interconnection system. However, AWE algorithm has the problem of numerical instability, which is mainly due to the power iteration of matrix in its moment calculation process. To solve the numerical stability problem of the explicit moment calculation of AWE algorithm, implicit moment matching methods based on Krylov subspace projection have been proposed successively. Typical algorithms include PVL based on Lanczos process\cite{ref32,ref33} and Arnoldi process\cite{ref34}. Then, aiming at the passivity problem of reduced order system, PRIMA based on Arnoldi process is proposed. It has many advantages, but it does not maintain the reciprocity of interconnected circuits well. In order to solve this problem, MOR based on structure preservation have been proposed successively\cite{ref35,ref36}. These methods are based on block space projection to maintain the block structure of the reduced order system, typically SPRIM algorithm \cite{ref37}, which is mainly based on the characteristics of MNA state matrix for block structure segmentation projection. At the same time, on the basis of first-order system order reduction, the second-order projection or moment matching has also been successfully developed, such as ENOR\cite{ref38} and SAPOR\cite{ref39}. In short, MOR based on moment matching mainly includes two parts. The first part is the state transformation. In the transformed state space, the state variables can be sorted according to the important characteristics of the circuit measurement. The second part is to cut off the least important state variables, so as to achieve the reduction of space dimension. 

\underline{Truncated Balanced Realization (TBR)}. TBR was first introduced by Moore\cite{ref40} to describe some states in a system that are difficult to reach and observe. The first-order Truncated Balanced methods applied in the control field include Lyapunov Balance method\cite{ref40} and random Balance method\cite{ref41}. In order to solve the problem of Lyapunov equation, a large number of Gramian approximation methods have been proposed\cite{ref42,ref43,ref44}, which mainly use the principal subspace method of approximating Gramian to speed up the solution of Gramian equations, such as PMTBR\cite{ref42}. The essence of the truncated balanced method is a kind of energy balance realization, so the balanced order reduction method should first make some realization of the original system, such as singular value realization and balanced realization, and then preserve some main characteristics to truncate. 

The most mature mainstream approach is the MOR based on moment matching. As our algorithm will utilize the results of moment matching algorithms, we describe the method presented in \cite{ref2} below.

Using the Modified Nodal Analysis (MNA) circuit state equation representation, a linear circuit with time as variable is expressed by the first-order differential equation as (\ref{eq4}).
\begin{equation}
	\label{eq4}
	\begin{cases}
		C\dot x_n=-Gx_n+Bu_N\\
		i_N=L^Tx_n
	\end{cases}
\end{equation}
where vector $x_n\in \mathbb{R}^{n\times 1}$ represents the state variable composed of the node voltage, the inductance and the new current introduced by the current source. $G, C\in \mathbb{R}^{n\times n}$ are extracted from the parasitic parameters of the interconnection line, $G$ represents the contribution of conductance, and $C$ represents the contribution of capacitance and inductance. The $u_N$ and $i_N$ vectors denote the port voltages and currents. $B, L\in \mathbb{R}^{n\times N}$ are the input and output incidence matrices.

Let $A=-G^{-1}C$ and $R=G^{-1}B$, so the admittance matrix of the system becomes (\ref{eq5}). The utilization of sparse matrix solvers (such as Sparse LU) can significantly expedite this process.
\begin{equation}
	\label{eq5}
	Y(s)=L^T(I_n-sA)^{-1}R
\end{equation}

By employing the Block Arnoldi algorithm, (\ref{eq6}) is obtained.
\begin{equation}
	\label{eq6}
	\begin{cases}
		colspan(X)=Kr(A,R,q)\\
		X^TAX=H_q\\
		X^TX=I_q
	\end{cases}
\end{equation}

The function of Block Arnoldi algorithm is to convert system matrix $A$ into upper Hessenberg matrix $H_q$. Due to the specific form of the upper Hessenberg matrix, the matrix inversion process in equation (\ref{eq6}) becomes notably expeditious. Let $x_n = X\cdot z_q$, where $z_q\in \mathbb{R}^{q\times 1}$ is the variable of the reduced order system, give the equations (\ref{eq7}) and (\ref{eq8}).

\begin{equation}
	\label{eq7}
	\begin{cases}
		H_q\dot z_q=z_q-X^TRu_N\\
		i_N=L^TXz_q
	\end{cases}
\end{equation}

\begin{equation}
	\label{eq8}
	Y(s)=L^TX(I_q-sH_q)^{-1}X^TR
\end{equation}

Eigendecomposition is used to calculate the $y$-parameter of the reduced order system. After the poles and residues are found, it can be expressed as (\ref{eq9}), where $q$ is the number of moments to match in the reduced RC model. It is necessary to choose a suitable matching order between accuracy and speed, \cite{ref45} select the order of the driving point model based on bandwidth estimation.

\begin{equation}
	\label{eq9}
	Y(s)=\sum_{i=j}^{q}\frac{res_j}{1-\frac{s}{pole_j}}
\end{equation}

\begin{figure}[!t]
	\centering
	\begin{tabular}{c}
		\includegraphics[width=3in]{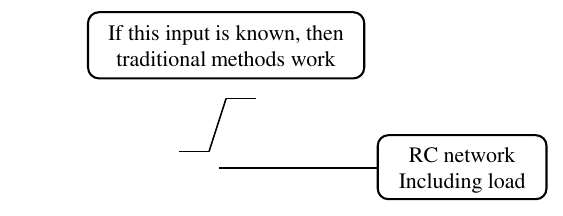}\\
		(a)\\
		\includegraphics[width=3in]{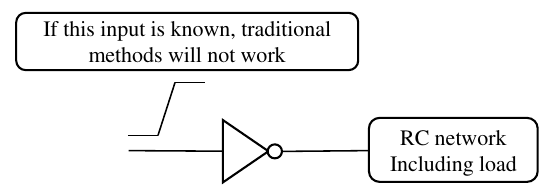}\\
		(b)\\
	\end{tabular}
	\caption{Limitations of traditional algorithms for RC modeling of signal lines.}
	\label{Fig.1}
\end{figure}

In short, with the application of moment matching techniques, the driving point function of RC load can always be represented in the form of (\ref{eq9}). A higher-order driving point function provides results closer to distributed RC than the $\pi$-model. When the input to the RC network is known, accurate current/voltage response predictions can be obtained using existing algorithms, as illustrated in Fig. \ref{Fig.1} a.

The inputs of a signal line are not always constant and their waveforms strongly depend on both the characteristics of the driver and the load. In response to any change in the driver or the RC load, the current will immediately change, making the algorithms like PRIMA less effective since the current waveform at the driver's output which is the input of the RC network is unknown and needs to be simulated, as depicted in Fig. \ref{Fig.1} b.



\section{DCM-BASED CURRENT WAVEFORM ESTIMATION ALGORITHM}
\subsection{Basic Assumptions and Theoretical Basis for the Proposed Algorithm}
Since the signal line current waveform involves both the driver and the load, it is less effective to use traditional approaches to predict the exact current amplitude without performing transistor level simulation. We propose a novel approach for computing the current response of signal line RC networks, enabling precise prediction of voltage and current waveforms.

We notice that all traditional algorithms compute the current response by either simplifying the driver model or the load model. But, regardless of which part is simplified, it is impossible to maintain a consistently accurate current response. Deviations in waveform in one aspect can lead to incorrect conclusions. An ideal approach would require both the driver and load characteristics to be taken into consideration. Recall that moment matching algorithms can efficiently compute the RC current response as long as the input waveform information is given; the challenge becomes how to obtain the input waveform of the RC load, which is the output of a driver. Indeed, without performing transistor level simulation, it is unlikely to capture the exact current waveform. Some additional information is required to bypass the transistor level simulation. We take advantage of the driver characterization data that is commonly supplied by the foundry.

We develop our method based on the assumption that current (or voltage) responses to the known inputs applied to the driver loaded with different purely capacitive loads are known. Foundries commonly provide them. In case they are not available, SPICE-based pre-characterization using fixed capacitors is required. As long as this information is given, the driver's current response can be determined. This can be illustrated using the current capacitance relation. For any RC load driven by a driver, the current can always be computed using the following (\ref{eq10}).

\begin{equation}
	\label{eq10}
	I=C\cdot \frac{dV}{dt}
\end{equation}
\begin{equation}
	\label{eq11}
	I(t_n)=\sum_{i=1}^{n}C\cdot \frac{dV_i}{dt_i}
\end{equation}

For the same driver, if the environmental parameters ($V_{dd}$, temperature, etc.) do not change, its output current and voltage value remain identical for the same capacitance load at the same time. This means that the actual current response of a complicated RC load can be seen as a superposition of current responses of multiple pure capacitances in different time frames, as expressed by (\ref{eq11}).

To demonstrate the driving capability and to perform delay calculation, foundries provide the current response information as a time versus current amplitude table for each driver loaded with specific capacitive loads. Furthermore, interpolation (or/and extension) fitting functions are employed to describe the transient behavior. 

The characterization library saves a small number of driving characteristic curves under various Vdd and temperature, allowing for the interpolation of pre-characterized waveforms at runtime according to the set environment. This ensures that the waveforms produced are highly precise and tailored to the specific design requirements.

If the driver characteristic data provided by the foundry are not available, we can obtain the necessary data performing SPICE simulations. But, this driver RC response characterization must be done in limited simulation runtime. Otherwise, performing a direct transistor level SPICE simulation on the target RC network would be much easier and more convenient. In order to limit the total runtime to characterize the RC response, we set a finite range of the capacitance load and specify its resolution. For example, for a specific driver, we approximate its maximum load as the sum of lump capacitances in its own RC network, recorded as $C_{max}$. The resolution of 5\% of $C_{max}$ is accurate enough for RC characterization and limits the total number of simulations to 20. To characterize those capacitance values below the 5\% of $C_{max}$ or between the two reference values, an interpolation technique is used.

\subsection{The Algorithm and its Implementation}
\begin{figure}[!t]
	\centering
	\includegraphics[width=3in]{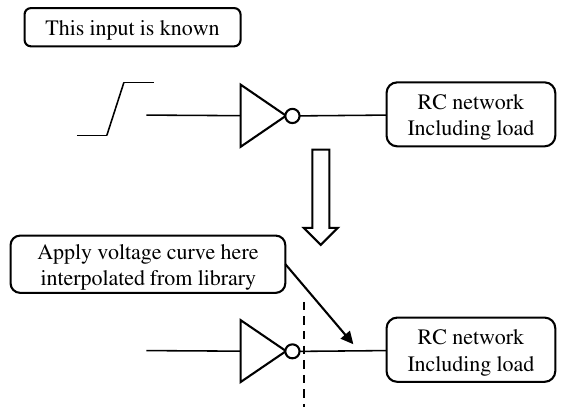}
	\caption{High level view of the proposed approach. Replace the driver and its input with the voltage curve interpolated from the driver characterization library.}
	\label{Fig.2}
\end{figure}

The approach we propose here replaces the original driver and input by the voltage curve interpolated (or/and extended) from the driver response table, and then calculates and validates the response using the existing data in the library, as shown in Fig. \ref{Fig.2}.

In industry, current responses for a specific driver are typically provided as a 3D table showing the current response value versus time and load capacitance.

\begin{figure}[!t]
	\centering
	\includegraphics[width=3in]{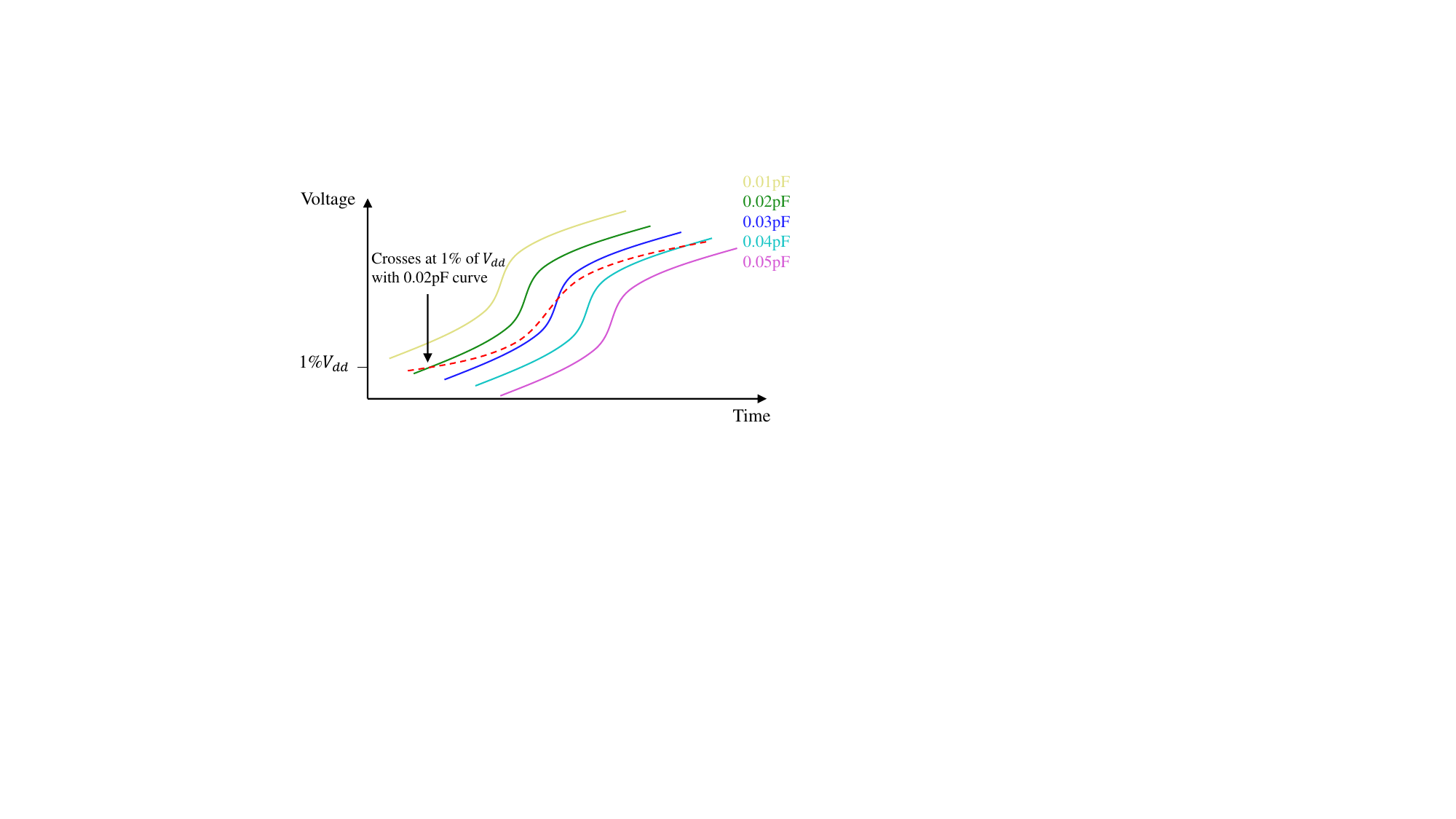}
	\caption{Voltage versus time plane. The black curves are interpolated from data library as the output voltage responses for different capacitance values of a specific driver (Eg: 0.01 pF to 0.05 pF). The red dashed line is the actual voltage response of an RC circuit crossing the black lines at different points.}
	\label{Fig.3}
\end{figure}

When we interpolate the discrete current table data as a piecewise function and integrate into voltage, we obtain a voltage response versus time function $V(t)$ for different capacitive loads for the specific driver, as illustrated in Fig. \ref{Fig.3}, it shows a voltage versus time plane for the driver with modified capacitive load. Different loads result in different voltage response curves in this plane. The output voltage curve of an RC circuit, which is a function of time, is also plotted on this plane. However, the actual curve will be completely different than any curve that maintains a fixed capacitance. Since the effective capacitance of an RC circuit changes over time, the actual curve crosses multiple curves of fixed capacitance until the effective capacitance saturates as the voltage reaches $V_{dd}$. Whenever the output voltage curve crosses one of the capacitance curves, the time point and output voltage of these two curves are identical. At that time, the capacitance value is the effective capacitance of this RC circuit and the voltage value is the same as the actual voltage of this RC circuit. Therefore, the output voltage of any RC network can be treated as a curve that is obtained by connecting points on different curves of single capacitance responses. In other words, we are constantly changing the value of the effective capacitance, its value is related to time, the driver and the RC network, showing a dynamic in the process.

In order to predict the output voltage without conducting a simulation, we divide the output curve into several regions and determine the average capacitance load associated with each region. Since the driver characterization data is discrete, the average capacitance may not be an exact value in the table. To generate the required $I(t)$ or $V(t)$ for the given capacitive load, interpolation or extrapolation is performed. Our experiments have shown that linear interpolation is accurate enough as long as the resolution of the original table is relatively high (Eg: 5\% of $C_{max}$).

For example, set the number of algorithm segments $N$=100, that is, select 100 equally spaced points. Then the 1\% of $V_{dd}$ point of the actual RC response will be on one of the curves (0.02 pF) in the plane with the same 1\% $V_{dd}$, as shown in Fig. \ref{Fig.3}. Let us assume the 1\% of $V_{dd}$ point belongs to the curve with capacitance value $C_{eff, 1\%}$ (the effective capacitance when the voltage reaches 1\% of $V_{dd}$, corresponding to 0.02 pF in Fig. \ref{Fig.3}), then the actual current of the RC circuit will be the same as the current of $C_{eff, 1\%}$ when their voltages are both at 1\% of $V_{dd}$. Thus as long as we can find the $C_{1\%}$ curve, we can find the current of the RC circuit when its voltage reaches 1\% of $V_{dd}$.

\begin{figure}[!t]
	\centering
	\includegraphics[width=3in]{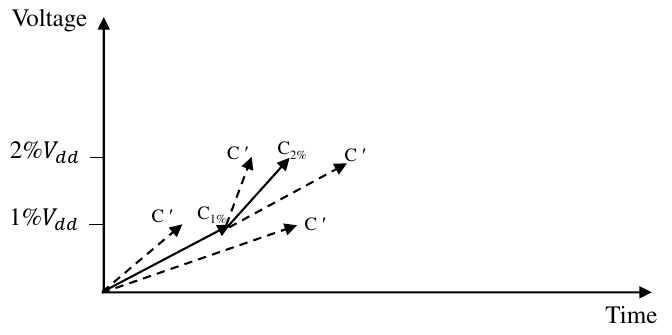}
	\caption{Searching the effective capacitance for each voltage step based on the driver characterization table.}
	\label{Fig.4}
\end{figure}

\begin{figure}[!t]
	\centering
	\includegraphics[width=3in]{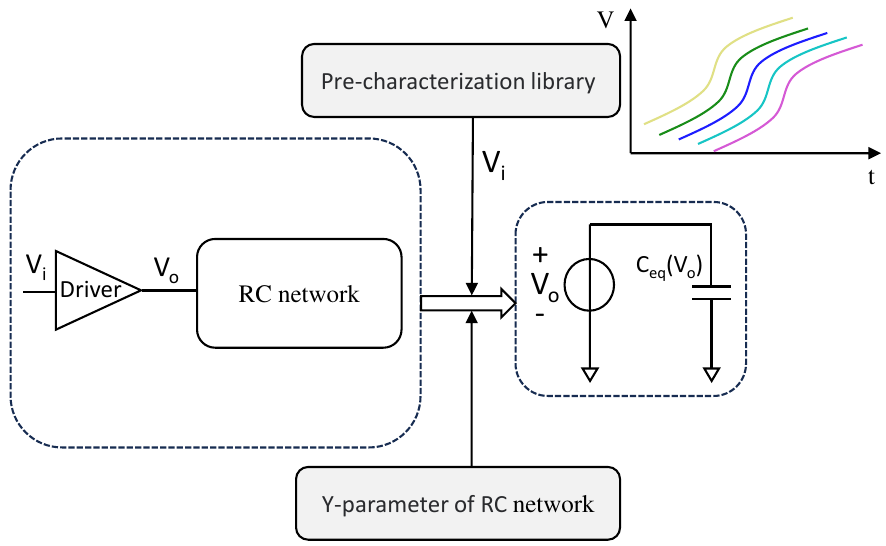}
	\caption{A (DCM)-based model.}
	\label{Fig.5}
\end{figure}

Starting from 0 to 1\% of $V_{dd}$, we choose one of capacitance curves ($C'$) and apply the voltage source to the RC network. We use an existing algorithm like PRIMA to speed up the calculation of current. Compare the calculated current at 1\% of $V_{dd}$ point and the current of $C$ from the table. If these two values match (difference less than tolerance error), we claim that $C$ is the $C_{1\%}$ that we seek. If these two current values do not match, we choose another capacitance curve and repeat this process until we find the correct one, as illustrated in Fig. \ref{Fig.4}. The DCM-based model is depicted in Fig. \ref{Fig.5}. It selects a pre-characterized library based on the driver's gate slew. The simplified model consists of a parallel combination of a voltage source and a capacitance. This capacitance, denoted as $C_{eq}$, is a function of the output voltage, varying among N values, representing the effective capacitance at different points. The more points we select, the more accurate the results become. We will discuss the matching strategy later in the paper.

The key issue is how to quickly and accurately obtain the response current value of the RC network during the matching process, so as to compare it with the library current. We introduce the moment matching algorithm, which simplifies the complex RC network system and obtains the impedance function $Z(s)$ or admittance function $Y(s)$, which is consisting of residue-pole pair. There are two traditional methods to obtain the time-domain current response based on the $y$-parameter. One is the inverse Laplace transform, as shown in (\ref{eq12}), which is time-consuming and has numerical errors. Another way is the convolution integral, as shown in (\ref{eq13}), which requires $O(T^2)$ complexity, $T$ is the number of time points during simulation.

\begin{equation}
	\label{eq12}
	i(t)=\mathcal{L}^{-1}[U(s)\cdot Y(s)]
\end{equation}
\begin{equation}
	\label{eq13}
	i(t)=\int_{0}^{t}y(t-\tau)v(\tau)d\tau
\end{equation}

These are numerical operations, and only approximate values can be obtained, especially when the RC network is large in scale or small in value. Therefore, we use symbolic expressions to speed things up and eliminate cumulative errors.

\begin{figure}[!t]
	\centering
	\includegraphics[width=3in]{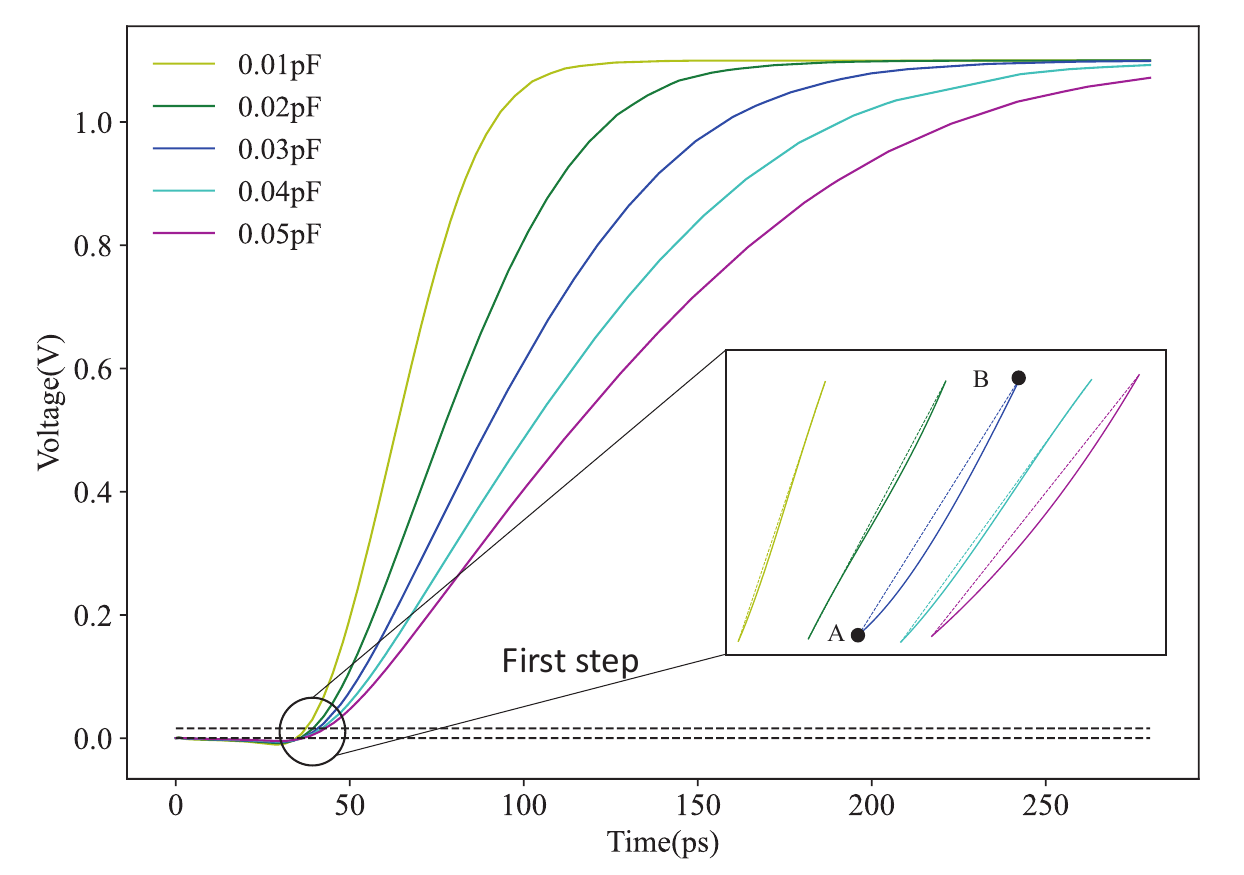}
	\caption{Driver voltage characterization table. The voltage response curves of different load capacitors are approximated by a straight line if the resolution is sufficient.}
	\label{Fig.6}
\end{figure}

When we set the resolution enough (Eg: 1\%$V_{dd}$), its voltage excitation can be replaced by a straight line to obtain approximate current response, similar to the limit, as shown in Fig. \ref{Fig.6}. The linear function determined at two points in the voltage pre-characterization curve can be used as part of the piecewise voltage excitation, as shown at points A and B (for 0-1\%$V_{dd}$). We use the piecewise linear (PWL) as the input source of the RC network, according to the fixed form of the $y$-parameter obtained by the moment matching algorithm, we get a symbolic expression, that is, we only need to perform algebraic operations to get the corresponding current response.

\begin{equation}
	\label{eq14}
	v(t)=\sum_{i=1}^{N-1}(k_i(t-t_i)+v_i)u(t-t_i)u(t_{i+1}-t)
\end{equation}

The voltage PWL is given by (\ref{eq14}), where $k_i=\frac{v_{i+1}-v_{i}}{t_{i+1}-t_{i}}$, $u(t)$ is the Heaviside step function, and $N$ is the number of segments set by the algorithm, it determines the number of steps (matching points) we need to generate the driver output response. Determine the expression of N-1 segment by continuous iteration, as shown in Fig. \ref{Fig.4}. The frequency domain expression can be further obtained as (\ref{eq15}).

\begin{eqnarray}
	\label{eq15}
	V(s)&=&\mathcal{L}\{v(t)\} \nonumber \\
	&=&\int_{0}^{+\infty}v(t)e^{-st}dt \nonumber \\
	&=&\sum_{i=1}^{N-1}\int_{t_{i}}^{t_{i+1}}(k_i(t-t_i)+v_i)e^{-st}dt \nonumber \\
	&=&\sum_{i=1}^{N-1}((k_i\frac{1}{s^2}+v_i\frac{1}{s})e^{-st_i} \nonumber \\
	&\;&-(k_i\frac{1}{s^2}+v_{i+1}\frac{1}{s})e^{-st_{i+1}})
\end{eqnarray}

Taking the inverse Laplace transform of (\ref{eq9}) and (\ref{eq15}) yields (\ref{eq16}), shown at the bottom of the next page. For a certain residue-pole pair, adding each segment in VPWL gives the partial values, and then the current can be obtained by adding each partial value. After the symbolic expression (\ref{eq16}) is obtained through this analysis, the input excitation coordinates $(t_i, v_i$) and the res-pole corresponding to the RC network can be substituted to obtain $i(t)$ efficiently and accurately. At the beginning of the algorithm, (\ref{eq14}) has only one section, and the number of sections is increased until N-1 by finding the matching dynamic capacitance value in iterations, that is, $v(t)$ contains the previously matched excitation section and is used for the calculation of the next section. After the iteration is completed, (\ref{eq14}) and (\ref{eq16}) are used as the voltage/current waveform of the RC network.

\begin{algorithm}[!t]
	\caption{DCM-based RC Current Response Computation.}\label{alg:alg1}
	\renewcommand{\algorithmicrequire}{\textbf{Input:}}
	\renewcommand{\algorithmicensure}{\textbf{Output:}}
	\begin{algorithmic}[1]
		\REQUIRE Drivers pre-characterized library; RC load network and the name of its driver.
		\ENSURE  Response of each driver(voltage/current); AVG, RMS and Peak current.
		
		\FOR{each combination of driver and load network}{
			\STATE Identify the driver's table;
			\STATE Set the parameter $N$;
			\STATE Calculate $y$-parameter of RC network;
			\FOR{each step}{
				\STATE Select the intermediate capacitance in the table to start binary search, interpolate and get coordinates $(t_i,v_i),(t_i,i_i)$;
				\STATE Append the matching results of each previous step to construct the PWL of $v(t)$
				\STATE Calculate $i(t_i)$ according to \ref{eq16};
				\STATE Compare this value to the table data $i_i$;
				\IF{two current values match} {
					\STATE Record the voltage/current as the actual value of this step;
				}
				\ELSE{
					\STATE Replace the matching capacitance of this step according to the search strategy, go back to 7;
				}
				\ENDIF
			}
			\ENDFOR
			\STATE Capture its over/under-shoot and tail from the pre-characterized library according to $C_{eff, 1}$ and $C_{eff, N-1}$ respectively;
			\STATE Fit the record points for each step;
			\STATE Calculate its AVG, RMS and Peak current;
		}\ENDFOR
	\end{algorithmic}
	\label{alg1}
\end{algorithm}

In the search strategy, Since the slope of the voltage response of different load capacitors is monotonic in any step, and there is a unique value $C_{eff}$ in each step. According to (\ref{eq10}), when the current value obtained in (\ref{eq16}) is smaller than the $i(t)$ in pre-characterized library, slope should be increased, that is, search to the left (decrease capacitance); Otherwise, we need to decrease slope and search to the right. Binary search combined with this strategy can effectively reduce the matching time.

\begin{figure}[!t]
	\centering
	\includegraphics[width=2.5in]{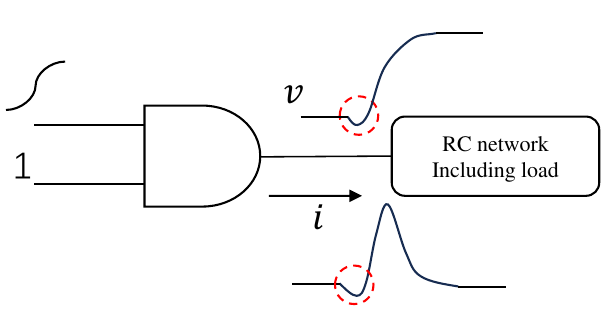}
	\caption{A rapidly changing input signal induces a reverse current at the output.}
	\label{Fig.7}
\end{figure}

A rapidly rising or falling input waveform will couple to the signal node, causing a reverse current at the output, and the node voltage rises above $V_{dd}$ or below $V_{ss}$, as shown in Fig. \ref{Fig.7}, the voltage or current is initially overshooted and/or undershooted due to the internal capacitance of driver (such as the overlay/channel capacitance of the MOSFET). This part of the response curve is not monotonous, and it is difficult to solve it by algorithm. In order to quickly obtain this part of the response, the method we adopt is to directly grab the corresponding over/under-shoot curves from the pre-characterized library according to the $C_{eff, 1\%}$. The reason for this is that the $C_{eff}$ of the RC network changes gradually when the voltage of the signal line node changes, and can be approximated by the $C_{eff}$ of adjacent steps. Furthermore, in order to ensure convergence at the step of 99\%$V_{dd}$ to $V_{dd}$, the curve corresponding to $C_{eff, 99\%}$ in library is connected to the tail of the prediction result.
\begin{figure*}[!t]
	\centering
	\includegraphics[width=6.5in]{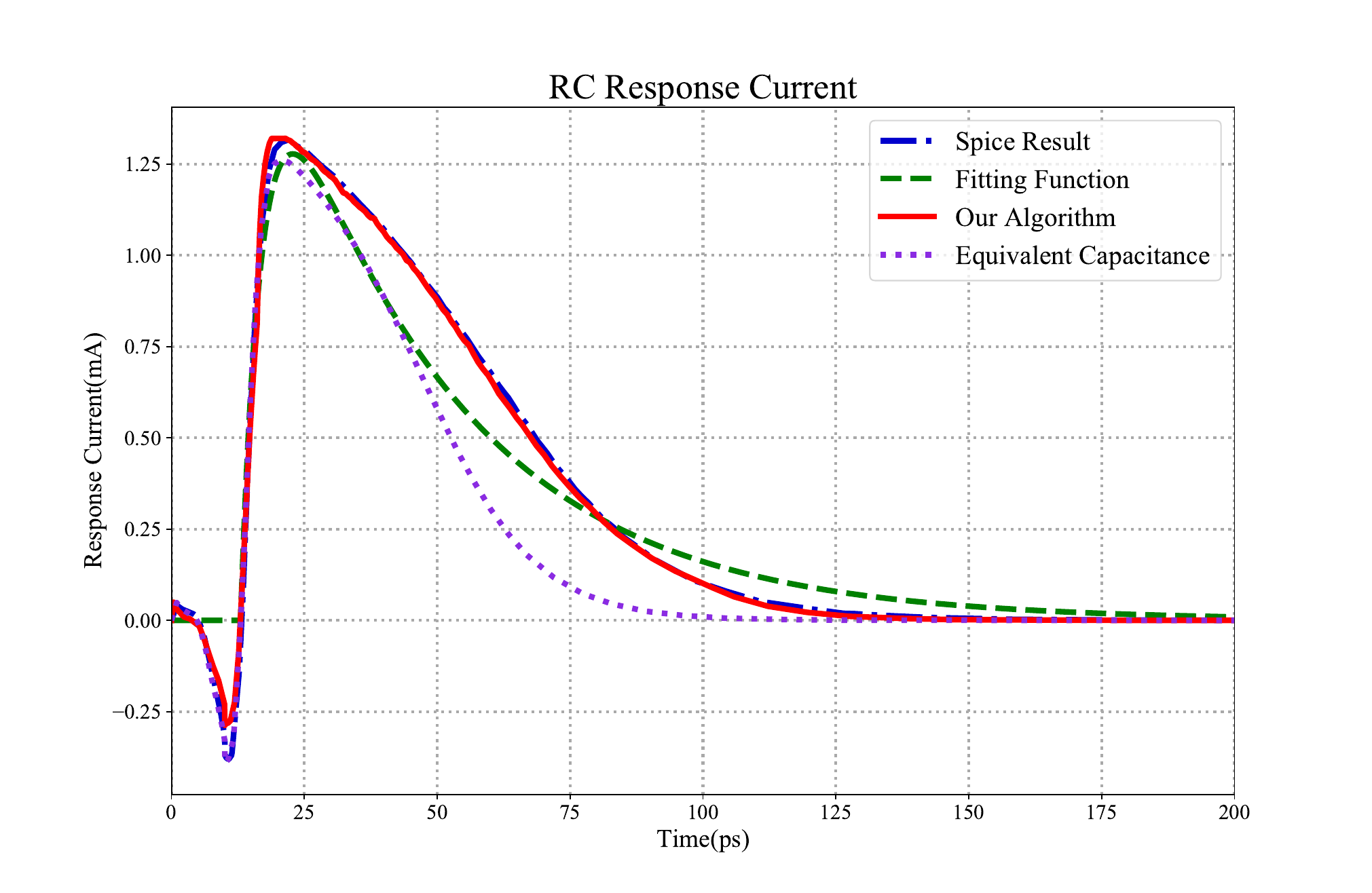}
	\caption{Current waveforms for the proposed algorithm and previous methods for the test benchmark.}
	\label{Fig.8}
\end{figure*}
The whole process is summarized in algorithm \ref{alg1}.
\begin{figure*}[!b]
	\hrulefill
	\begin{eqnarray}
		\label{eq16}
		i(t)&=&\mathcal{L^{\text{-1}}}\{V(s)Y(s)\} \nonumber \\
		&=&\sum_{j=1}^{q}\sum_{i=1}^{N-1}\frac{res_jk_i}{pole_j}((pole_j(t-t_i)-e^{(t-t_i)pole_j}+1)u(t-t_i)-(pole_j(t-t_{i+1})-e^{(t-t_{i+1})pole_j}+1)u(t-t_{i+1}))	\nonumber \\
		&\;&+\sum_{j=1}^{q}\sum_{i=1}^{N-1}res_j(v_i(1-e^{t-t_i})u(t-t_i)-v_{i+1}(1-e^{t-t_{i+1}})u(t-t_{i+1}))	\nonumber \\
		&=&\sum_{j=1}^{q}\sum_{i=1}^{N-1}\frac{res_jk_i}{pole_j}((pole_j(t-t_i)-e^{(t-t_i)pole_j}+1)u(t-t_i)-(pole_j(t-t_{i+1})-e^{(t-t_{i+1})pole_j}+1)u(t-t_{i+1}))	\nonumber \\
		&\;&+\sum_{j=1}^{q}res_j(v_1(1-e^{t-t_1})u(t-t_1)-v_{N}(1-e^{t-t_{N}})u(t-t_{N}))
	\end{eqnarray}
\end{figure*}

\section{SIMULATION RESULTS}

To prove the accuracy of the algorithm, we compared the proposed algorithm with previous work. A large range of load conditions and input slope are simulated and the influence of the parameter $N$ is analyzed. Furthermore, we evaluated the runtime of a complete project, including an analysis of several typical signal line benchmarks. We compared the simulation results with traditional methods and SPICE golden results. We assessed the fitting quality of the current waveform using AVG, RMS, and Peak errors. The experiments were conducted using a complete foundry model of a 40nm CMOS process with a 1.1 V power supply.

In Fig. \ref{Fig.8}, we show the current waveform for a typical clock line benchmark with roughly 2000 resistors and capacitors. The complex RC load of this network has obvious resistance shielding, so the gate output has a large exponential tail. It can be seen that our method achieves very good performance in both current and voltage prediction, well predicting the undershoot waveform in the initial part, leading to a 0.9\% error in AVG (1/2 cycle), 0.6\% error in RMS and 0.3\% error in peak current for this circuit. The total run time is less than 1 second excluding the pre-characterization of the driver.

\begin{table}[!t]
	\begin{center}
		\caption{The intermediate steps of computing RC current response for the test benchmark.}
		\label{tab1}
		\begin{tabular}{cccc}
			\hline
			\textbf{Time(ps)} & Voltage(V) & Current(mA) & \textup{$C_{step}(fF)$}\\
			\hline
			\textbf{0}& 0 & 0 & 0\\
			\hline
			\textbf{10.9}&-0.026&-0.27&13\\
			\hline
			\textbf{16.1}&0.016&0.88&13\\
			\hline
			\textbf{16.6}&0.038&1.07&20\\
			\hline 
			\textbf{17.3}&0.059&1.21&28\\
			\hline
			\textbf{18.8}&0.102&1.32&36\\
			\hline
			\textbf{29.8}&0.361&1.22&39\\
			\hline
			\textbf{36.9}&0.512&1.11&41\\
			\hline
			\textbf{44.8}&0.650&0.97&43\\
			\hline
			\textbf{71.5}&0.964&0.43&44\\
			\hline
			\textbf{100.7}&1.072&0.11&45\\
			\hline
		\end{tabular}
	\end{center}
\end{table}

Table \ref{tab1} gives the dynamic capacitance value of each step for the above benchmark during the current estimation. Due to the resistive shielding effect, the effective capacitance is very small at the beginning and gradually saturates until all the capacitances are fully charged.

\begin{figure}[!t]
	\centering
	\includegraphics[width=3.3in]{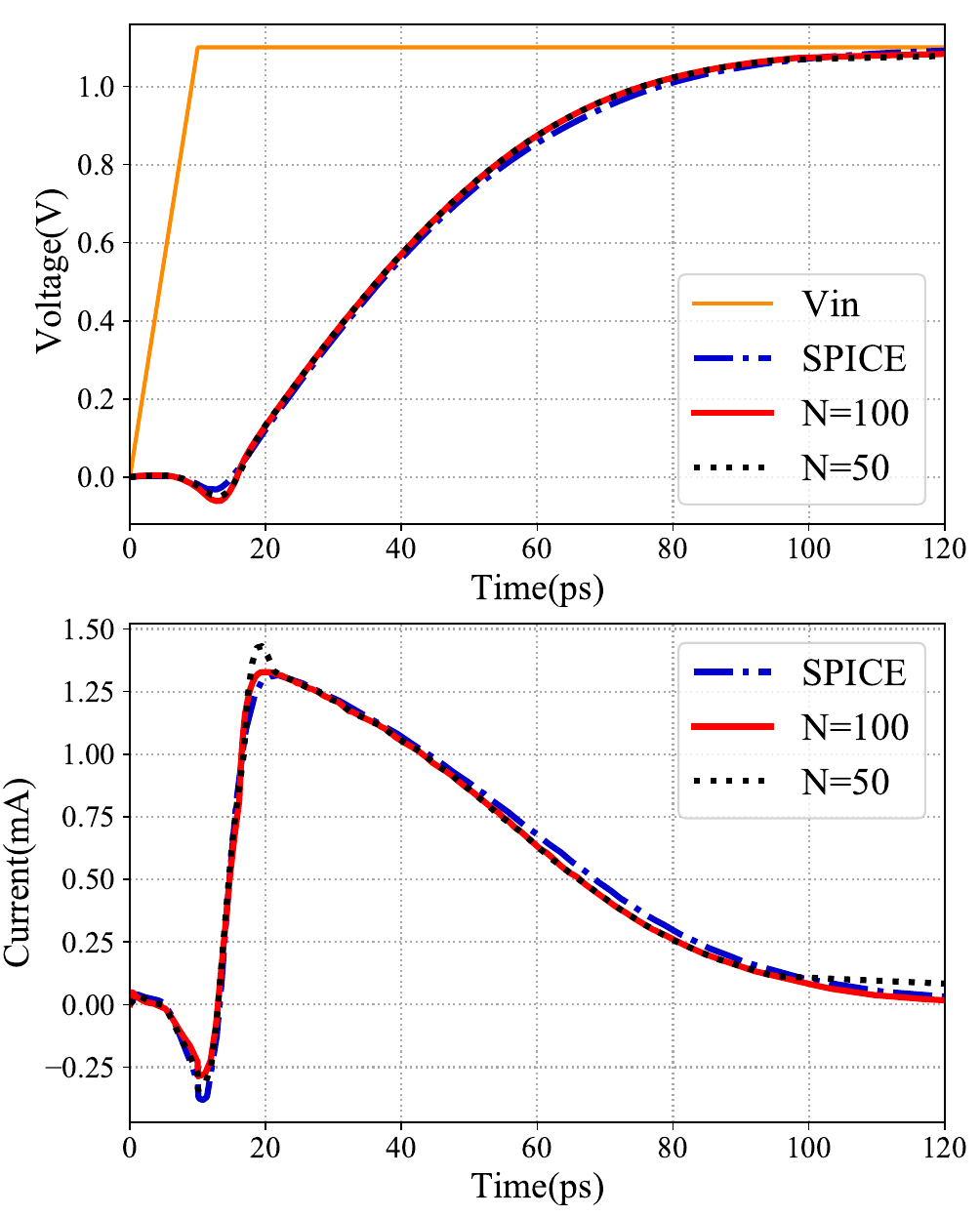}
	\caption{Apply different $N$ to analyze a steep ramp input.}
	\label{Fig.9}
\end{figure}

\begin{figure}[!t]
	\centering
	\includegraphics[width=3.1in]{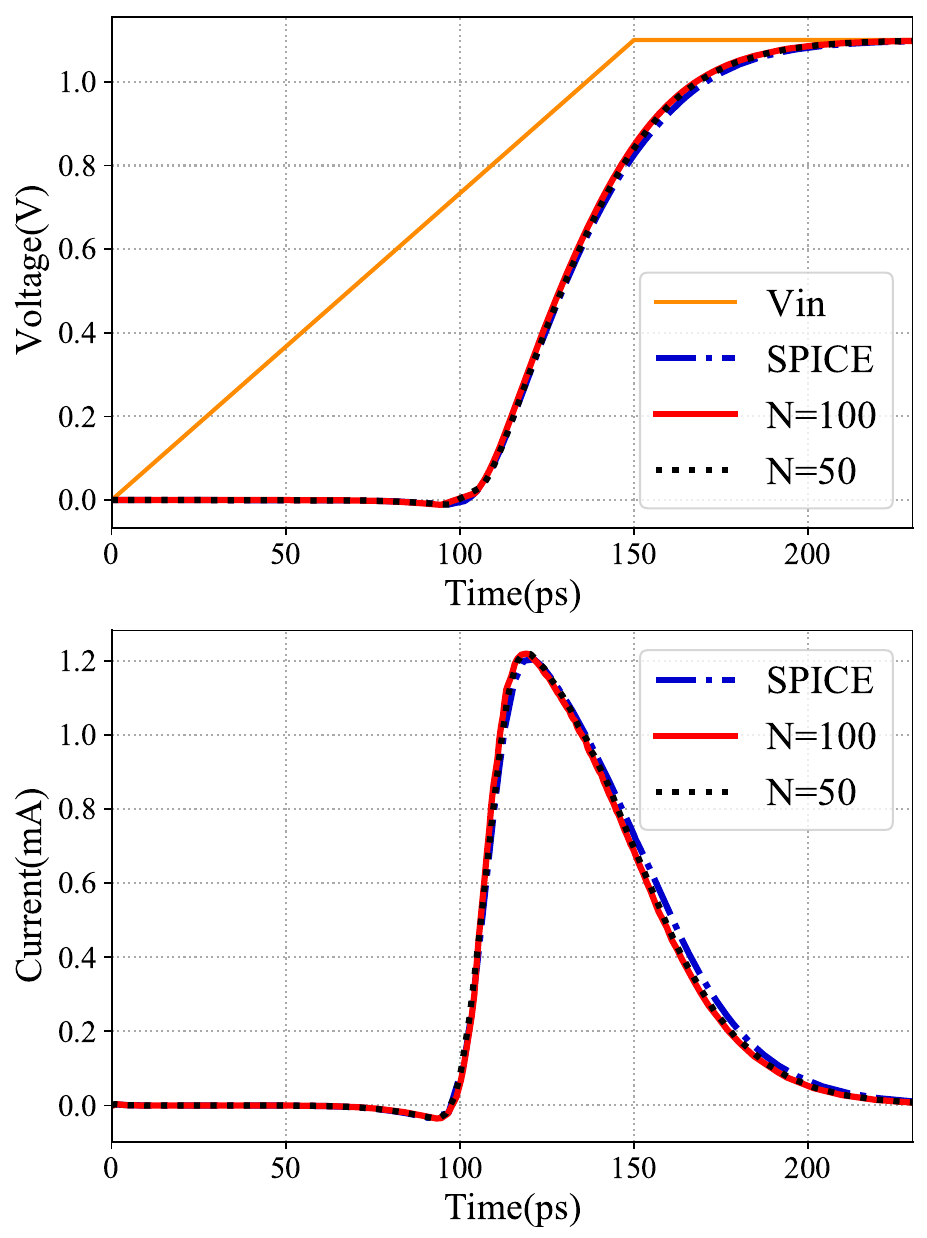}
	\caption{Apply different $N$ to analyze a normal input.}
	\label{Fig.10}
\end{figure}

So as to detect the impact of the number of iteration steps N, we set up different $N$ for analyses. When the input signal changes very quickly (0.01 ns), it will have a negative impact due to insufficient points, as shown in Fig. \ref{Fig.9}. When $N$ changes from 100 to 70 (not shown) and 50, the current error increases by 0.68\% and 4.74\% respectively. Fig. \ref{Fig.10} shows that for a normal transition time (0.15 ns), $N$=50 is accurate enough. For high-speed circuits with advanced processes, sufficient calculation points ensure accurate analysis results, while analysis points can be reduced for ordinary circuits or non-signoff standards. Calculate according to usage needs.

\begin{table}[!t]
	\begin{center}
		\caption{The program running of the proposed algorithm under different calculation methods (N=100).\\Equipped with AMD EPYC 64-core Processor @ 2.0GHz.}
		\label{tab2}
		\begin{tabular}{cccc}
			\hline
			Scale & Method & Runtime & Nets per second\\
			\hline
			\multirow{3}*{24091}& Laplace & 558 s & 43 \\
			\cline{2-4}
			~&Symbolic&\multirow{2}*{188 s}&\multirow{2}*{128}\\
			~&expression&~&~\\
			\hline
		\end{tabular}
	\end{center}
\end{table}

In Table \ref{tab2}, we analyzed a small 40nm CMOS chip containing tens of thousands of signal nets, each consisting of a standard cell and its load RC network. It can be seen that the program using \ref{eq16} is 3X faster than using the inverse Laplace transform. When $N$ is set to 50, the runtime is reduced by about 10\% because part of the program's time is spent solving $Y(s)$.

In order to further validate our algorithm, we simulated several typical representatives of signal line circuits including MUX (load as single driver), Bus-line (simple RC tree with multiple drivers modeled as load) with branches, and SRAM word-line (multiple driver and multiple load drivers). These signal line circuit models contain tens to thousands RC segment and cover the general topology of RC network that signal line is usually modeled with. The results including accuracy and runtime were compared with the NGSPICE circuit simulator results \cite{ref46}, as shown in Table \ref{tab3}. The reported runtimes do not include the driver pre-characterization step. If the driver characterization data provided by the foundry are not available, performing such a pre-characterization requires less than 1 minute per driver for 5\% of $C_{max}$ resolution. As shown in the table, the proposed algorithm can achieve excellent estimations of current waveforms and the target current values (AVG, RMS and peak) with almost 50-200X faster runtime than NGSPICE on typical signal benchmarks (5-20X faster if we need to pre-characterize driver).

\begin{table}[!t]
	\begin{center}
		\caption{Error of AVG, RMS and peak currents and runtime comparisons (N=100).}
		\label{tab3}
		\begin{tabular}{ccccc}
			\hline
			& AVG & RMS & Peak & Runtime\\
			& error & error & error &  \\ 
			\hline
			CLK(SPICE)& -- & -- & -- &80 s\\
			\cline{2-5}
			($C_{eff}$ model)&20\%&10\%&2\%& $<$1 s\\
			\cline{2-5}
			(Fitting function)&6\%&6\%&2\%& $<$1 s\\
			\cline{2-5}
			{\bf (Proposed model)}&{\bf 0.3\%}&{\bf 0.1\%}&{\bf 0.3\%}&{\bf $<$1 s}\\
			\hline
			BUS(SPICE)&--&--&--&40 s\\
			\cline{2-5}
			($C_{eff}$ model)&6\%&8\%&2\%& $<$1 s\\
			\cline{2-5}
			(Fitting function)&6\%&12\%&7\%& $<$1 s\\
			\cline{2-5}
			{\bf (Proposed model)}&{\bf 1.2\%}&{\bf 0.2\%}&{\bf 0.5\%}&{\bf $<$1 s}\\
			\hline
			MUX(SPICE)&--&--&--&16 s\\
			\cline{2-5}
			($C_{eff}$ model)&2\%&11\%&8\%& $<$1 s\\
			\cline{2-5}
			(Fitting function)&17\%&1\%&4\%& $<$1 s\\
			\cline{2-5}
			{\bf (Proposed model)}&{\bf 0.8\%}&{\bf 0.2\%}&{\bf 0.2\%}&{\bf $<$1 s}\\
			\hline
			SRAM(SPICE)&--&--&--&78 s\\
			\cline{2-5}
			($C_{eff}$ model)&9\%&15\%&4\%& $<$1 s\\
			\cline{2-5}
			(Fitting function)&18\%&3\%&11\%& $<$1 s\\
			\cline{2-5}
			{\bf (Proposed model)}&{\bf 0.2\%}&{\bf 0.3\%}&{\bf 0.2\%}&{\bf $<$1 s}\\
			\hline
		\end{tabular}
	\end{center}
\end{table}

It is worth noting that the proposed algorithm is highly scalable and can be applied to various signal line benchmarks with different levels of complexity, ranging from simple RC trees to complex bus structures and SRAM word-lines. These response waveforms serve as essential parameters for various metrics, including signal line EM reliability, timing analysis, and switching power, among others.

There is a small discrepancy between the results obtained by our algorithm and the golden results of SPICE simulation. The main contributors of the approximation error are the two steps of interpolation/extension (capacitance and voltage/current), Miller effect of the load and finite number of steps we choose. In this paper, we do not show the quantitative error composition. Besides this, there are also several natural limits for the proposed algorithm. First, if the driver is complicated (multi-stage) and the load is simple such that characterization of the driver would be time consuming rather than direct simulation. Second, part of the RC load consists of drivers of next stage that will have a significant Miller effect that cannot be modeled as fixed capacitance captured in the pre-characterization. Third, the load involves large capacitance so that pre-characterization would be either time consuming or less accurate due to the minimum resolution.

\section{CONCLUSION}
In this paper, we propose a dynamic capacitance matching (DCM)-based RC current response algorithm for calculating the current waveform of a signal line without the necessity of SPICE simulation is proposed. Unlike previous methods, our algorithm does not depend on function fitting or a single effective capacitance. Instead, it seamlessly integrates the high-order driving-point functions of the RC network with the pre-characterized library of drivers through symbolic expressions, swiftly performing algebraic manipulations to determine dynamic capacitance values under different output voltages. Dynamic capacitance precisely characterizes the behavior of gate cells under any RC load, exploiting their mutual interdependencies to precisely match the driver's output current and account for overshoots/undershoots at any given moment. This process is highly configurable, allowing for adjustments tailored to specific application scenarios. Experimental results show that the AVG, RMS and Peak current errors are nearly 1\% and the running time is 50-200X faster than the golden results obtained by NGSPICE.

\vfill

\end{document}